\documentclass[11pt]{iopart}
\usepackage{iopams}
\usepackage{txfonts}
\usepackage{graphicx}

\def\HII{H{\sc ii}}

\def\aap{Astron. Astrophys.}
\def\aaps{Astron. and Astrophys. Suppl. Sries}

\def\aj{Astron. J.}
\def\apj{Astrophys. J.}

\def\apjl{Astrophys. J. Letters}

\def\mnras{Mon. Not. R. Astron. Soc.}

\def\aapr{{Astron. Astrophys. Rev}}

\begin{document}

\article[Star Formation in \HII\ Galaxies]{Star-forming Dwarf Galaxies: Ariadne's Thread in the
  Cosmic Labyrinth}{Star Formation in \HII\ Galaxies. Properties of the ionized gas}

\author{G.~F.\ H\"agele$^{1}$, A.~I.\ D\'iaz$^{1}$, E.\ Terlevich$^{2}$, E.\
  P\'erez-Montero$^{3,4}$,\\ 
  R.\ Terlevich$^{2}$ and M.~V.\ Cardaci$^{1,5}$}

\address{$^{1}$ Departamento de F\'{\i}sica Te\'orica, C-XI, Universidad Aut\'onoma de
Madrid, 28049 Madrid, Spain\\
$^{2}$ INAOE, Tonantzintla, Apdo. Postal 51, 72000 Puebla, M\'exico\\ 
$^{3}$ Instituto de Astrof\'isica de Andaluc\'ia, CSIC, Apdo.\ 3004, 18080,
Granada, Spain\\
$^{4}$ Laboratoire d'Astrophysique de Toulouse-Tarbes. Observatoire
  Midi-Pyr\'en\'ees. 14, avenue Edouard Belin. 31400. Toulouse. France\\
$^{5}$ XMM Science Operations Centre, European Space Astronomy Centre of ESA,
P.O. Box 50727, 28080 Madrid, Spain\\
}
\ead{ghagele@uam.es}
\begin{abstract}

We present two set of high signal-to-noise spectrophotometric observations of
three and seven compact and luminous \HII\ galaxies observed with the WHT and
the 3.5\,m telescope at CAHA, respectively. All the observations have been
made with the use of a double-arm spectrograph which provides spectra with a 
wide wavelength coverage, from 3200 to 10500\,\AA for the WHT data and 3400 to
10400\,\AA for the CAHA data, of exactly the same region of a given
galaxy. These spectral ranges include the
[O{\sc ii}]\,$\lambda\lambda$\,3727,29\,\AA\ lines, the [S{\sc
    iii}]\,$\lambda\lambda$\,9069,9532\,\AA\ doublet as well as various weak
auroral lines such as [O{\sc iii}]\,$\lambda$\,4363\,\AA\ and [S{\sc
    iii}]\,$\lambda$\,6312\,\AA.

We propose a methodology to perform a self-consistent analysis of the
physical properties of the emitting gas of \HII\ galaxies adequate to the data
that can be obtained with the XXI century technology. This methodology
requires the production and calibration of empirical relations between the
different line temperatures that should superseed currently used ones based on
very simple, and poorly tested,  photo-ionization model sequences. 
Then, these observations are analysed 
applying a methodology designed to obtain accurate elemental abundances
of oxygen, sulphur, nitrogen, neon, argon and iron in the ionsied gas. 
Four electron temperatures and one electron density are derived 
from the observed forbidden line ratios using the five-level atom
approximation.  
For our best objects errors of  1\% in t$_e$([O{\sc iii}]),  
3\% in t$_e$([O{\sc ii}]) and 5\% in t$_e$([S{\sc iii}]) are achieved 
with a resulting accuracy between 5 and 9\% in total oxygen abundances,
O/H. These accuracies are expected to 
improve as better calibrations based on more precise measurements, both on 
electron temperatures and  densities, are produced.

For the objects observed with the WHT we have compared the measurements obtained
for our spectra with those performed on the spectra downloaded from the SDSS DR3
finding a satisfactory agreement.

The ionization structure of the nebulae can be mapped by the theoretical
oxygen and sulphur ionic ratios, on the one side,  and the corresponding
observed emission line ratios, on the other --  the $\eta$ and $\eta$' plots
--. The combination of both is shown to provide a means to test
photo-ionization model sequences currently applied to derive elemental
abundances in \HII\ galaxies. 
The ionization structure found for the observed objects from the
O$^{+}$/O$^{2+}$ and S$^{+}$/S$^{2+}$ ratios points to high values of the 
ionizing radiation, as traced by the values of the ``softness parameter" 
$\eta$ which is less than one for all the objects. The use of line
temperatures derived from  t$_e$([O{\sc iii}]) based on current photo-ionization
models yields for the two highest excitation objects much higher values of 
$\eta$ which would imply lower ionizing temperatures. This is however
inconsistent with the ionization structure as probed by the measured emission
line intensity ratios.

Finally, we have measured the T(Bac) for three of the \HII\ galaxies and
computed temperature fluctuations. Only for one of the objects, the
temperature  fluctuation is significant and could lead to higher oxygen abundances by about
0.20\,dex.

\end{abstract}

\maketitle

\section{Introduction}

\HII\ galaxies are low mass irregular galaxies with, at least, a recent episode
of  violent star formation \cite{1985MNRAS.216..255M,1985RMxAA..11...91M}
concentrated in a few parsecs close to their cores. The ionizing fluxes
originated by these young massive stars dominate the light of this subclass of
Blue Compact Dwarf galaxies (BCDs) which show  emission line spectra very
similar to those of giant extragalactic  \HII\ regions (GEHRs;
\cite{1970ApJ...162L.155S,1980ApJ...240...41F}). 
Therefore, by applying the same 
measurement techniques as for \HII\ regions, we can derive the temperatures,
densities and chemical composition of  the interstellar gas in this type of
generally metal-deficient galaxies
\cite{1991A&AS...91..285T,2000A&ARv..10....1K,2006MNRAS.365..454H}. In some
cases, it is possible to detect in 
these objects, intermediate-to-old stellar populations which have a more
uniform spatial distribution than the bright and young stellar  populations
associated with the ionizing clusters \cite{1998ApJ...493L..23S}. This older
population produces a characteristic spectrum with absorption features which
mainly affect the hydrogen recombination emission lines
\cite{1988MNRAS.231...57D}, that is the Balmer and Paschen series in the
spectral range of our interest. In some cases, the underlying stellar
absorptions can severely affect the ratios of H{\sc i} line pairs and hence
the determination of the reddening constant [c(H$\beta$)]. They must therefore
be measured with special care.

Spectrophotometry of bright \HII\ galaxies in the Local Universe allows the
determination of abundances from methods that rely on the measurement of
emission line intensities and atomic physics. This is referred to as the
"direct" method. In the case of more distant or intrinsically fainter
galaxies, the low signal-to-noise obtained with current telescopes precludes
the application of this method and empirical ones based on the strongest
emission lines are required. The fundamental basis of these empirical methods
is reasonably well understood (see e.g. \cite{2005MNRAS.361.1063P}).
The accuracy of the results however depends on the goodness of their calibration
which in turn depends on a well sampled set of precisely derived abundances by
the "direct" method so that interpolation procedures are reliable.

The precise derivation of elemental abundances however is not a
straightforward matter. Firstly, accurate measurements of the emission lines
are needed. Secondly, a certain knowledge of the ionization structure of the
region is required in order to derive ionic abundances of the different
elements and in some cases photo-ionization models are needed to correct for
unseen ionization states. An accurate diagnostic requires the measurement of
faint auroral lines covering a wide spectral range and their accurate (better
than 5\%) ratios to Balmer recombination lines. These faint lines are usually
about 1\% of the  H$\beta$ intensity. The spectral range must include from
the UV [O{\sc ii}]\,$\lambda\lambda$\,3727,29\,\AA\ doublet, to the near IR
[S{\sc iii}] 
$\lambda\lambda$\,9069,9532\,\AA\ lines. This allows the derivation of the
different line temperatures: T$_e$([O{\sc ii}]), T$_e$([S{\sc ii}]),
T$_e$([O{\sc iii}]), T$_e$([S{\sc iii}]), T$_e$([N{\sc ii}]), needed in order
to study the temperature and ionization structure of each \HII\ galaxy
considered as a multizone ionized region. 

Unfortunately most of the available starburst and \HII\ 
galaxy spectra have only a restricted wavelength  range (usually from about 3600
to 7000 \AA), consequence of observations with single arm spectrographs, and
do not have the adequate signal-to-noise ratio (S/N) to accurately measure the 
intensities of the weak diagnostic emission lines. Even the Sloan Digital Sky
Survey (SDSS) spectra \cite{2002AJ....123..485S} do not cover simultaneously the
[O{\sc ii}]\,$\lambda\lambda$\,3727,29 and the [S{\sc iii}]\,$\lambda$\,9069\,\AA\
lines, they only represent an average inside a 3\,arcsec fibre and reach the
required signal-to-noise ratio only for the brightest objects.
We have therefore undertaken a project with the aim of obtaining a database of
top quality line ratios for a sample that includes the best objects for the
task. The data is collected using exclusively two arm  spectrographs in order to
guarantee both high quality spectrophotometry in the whole spectral range from
3500 to 10500\,\AA\, approximately, and good spectral and spatial
resolution. In this way we 
are able to vastly improve constraints on the photo-ionization
models including the  mapping of the ionization structure and the measurement
of temperature fluctuations about which very little is known.

It is important to realize that the combination of accurate spectrophotometry
and wide spectral coverage cannot be achieved  using single arm spectrographs
where, in order to  reach the necessary spectral resolution, the wavelength
range must be split into several independent observations. 
In those cases, the
quality of the spectrophotometry is at best doubtful mainly because the
different spectral ranges are not observed simultaneously. 

\section{Observations and data reduction}

\subsection{WHT observations}

The blue and red spectra were obtained simultaneously using the Intermediate
dispersion Spectrograph and Imaging System (ISIS) double beam
spectrograph mounted on the 4.2m William Herschel Telescope (WHT) of the Isaac
Newton Group (ING) at the Roque de los Muchachos Observatory, on the Spanish
island of La Palma. They were acquired on July the 18th 2004 during one
single night observing run and under photometric conditions. EEV12 and Marconi2
detectors were attached to the blue and red arms of the spectrograph,
respectively. The R300B grating was used in the blue covering the wavelength
range 3200-5700\,\AA\ (centered at $\lambda_c$\,=\,4450\,\AA), giving a spectral
dispersion of 0.86\,\AA\,pixel$^{-1}$. On the red arm, the R158R grating was
mounted  providing a spectral range from 5500 to 10550\,\AA\
($\lambda_c$\,=\,8025\,\AA) and a spectral dispersion of
1.64\,\AA\,pixel$^{-1}$. The pixel size for this
set-up configuration is 0.2 arcsec for both spectral ranges.  The slit width
was $\sim$0.5 arcsec, which, combined with the spectral dispersions, yielded
spectral resolutions of about 2.5 and 4.8\,\AA\ FWHM in the blue and red arms 
respectively. All observations were made at paralactic angle to avoid effects of
differential refraction in the UV. 

The instrumental configuration was planned in order to cover the whole
spectrum from 3200 to 10550\,\AA\  providing at the same time a moderate
spectral resolution.   This 
guarantees the simultaneous measurement  of the Balmer discontinuity and the
nebular lines of [O{\sc ii}]\,$\lambda\lambda$\,3727,29 and [S{\sc
    iii}]\,$\lambda\lambda$\,9069,9532\,\AA\  at both ends of the spectrum, in
the very same region of the galaxy. A good signal-to-noise ratio was also
required to allow the detection and measurement of weak lines such as  [O{\sc
    iii}]\,$\lambda$\,4363, [S{\sc ii}]\,$\lambda\lambda$\,4068, 6717 and 6731,
and [S{\sc iii}]\,$\lambda$\,6312.

\subsection{CAHA observations}
Blue and red spectra were obtained simultaneously using the  double beam
Cassegrain Twin Spectrograph (TWIN) mounted on the 3.5m telescope of the Calar
Alto Observatory at the Complejo Astron\'omico Hispano Alem\'an (CAHA),
Spain. They were acquired in June 2006, during a four night observing run and
under excellent seeing and photometric conditions. Site\#22b and Site\#20b,
2000\,$\times$\,800\,px 15\,$\mu$m, detectors were attached to the blue and
red arms of the spectrograph, respectively. The T12 grating was used in the
blue covering the wavelength range 3400-5700\,\AA\ (centered at
$\lambda_c$\,=\,4550\,\AA), giving a spectral dispersion of
1.09\,\AA\,pixel$^{-1}$ (R\,$\simeq$\,4170). On the red arm, the T11 grating
was mounted  providing a spectral range from 5800 to 10400\,\AA\
($\lambda_c$\,=\,8100\,\AA) and a spectral dispersion of
2.42\,\AA\,pixel$^{-1}$ (R\,$\simeq$\,3350).  
The pixel size for this set-up configuration is 0.56\,arcsec for both spectral
ranges.  The slit width was $\sim$1.2\,arcsec, which, combined with the
spectral dispersions, yielded spectral resolutions of about 3.2 and 7.0\,\AA\
FWHM in the blue and the red respectively. 

Again, all observations were made at
paralactic angle to avoid effects of differential refraction in the UV. The
instrumental configuration covers the whole 
spectrum from 3400 to 10400\,\AA\ (with a gap between 5700 and 5800\,\AA)
providing at the same time a moderate spectral resolution.

\subsection{Data reduction}
\label{data}

All the images were processed and analysed with
IRAF\footnote{IRAF: the Image Reduction and Analysis Facility is distributed by
  the National Optical Astronomy Observatories, which is operated by the
  Association of Universities for Research in Astronomy, Inc. (AURA) under
  cooperative agreement with the National Science Foundation (NSF).} routines in 
the usual manner. The procedure includes
the removal of cosmic rays, bias subtraction, division by a normalized flat
field and wavelength calibration. 

\medskip

For a complete discussion about instrumental configuration, data reduction
procedure, results, method of abundance determinations and ionization
structure see \cite{2006MNRAS.372..293H,2008MNRAS.383..209H}.  

\section{Summary and conclusions}
\label{summ2}

We have performed a detailed analysis of newly obtained spectra of three and
seven \HII\ galaxies observed with the 4.2\,m WHT and the 3.5\, CAHA
telescopes, 
respectively. These galaxies were selected from the Sloan Digital Sky Survey
Data Release 2 and 3, respectively. For the first set of galaxies the spectra
cover from 3200 to 10550\,\AA\ in wavelength, while for the second group the
data cover from 3400 to 10400 \AA\ with a gap of approximately 100\,\AA\
between 5700 and 5800\,\AA. The WHT spectra have a FWHM resolution of 
about 1800 in the blue and 1700 in the red spectral regions, and the CAHA ones
of about 1400 and 1200, respectively. 
 
The high signal-to-noise ratio of the obtained spectra allows the measurement
of four line electron temperatures: T$_e$([O{\sc iii}]), T$_e$([S{\sc iii}]),
T$_e$([O{\sc ii}]) and T$_e$([S{\sc ii}]), for all the objects of the sample
with the addition of T$_e$([N{\sc ii}]) for three of them, and the Balmer
continuum temperature T(Bac) for the WHT objects. These
measurements and a careful and realistic treatment of the observational errors
yield total oxygen abundances with accuracies between 5 and 12\%. The
fractional error is as low as 1\% for the ionic O$^{2+}$/H$^{+}$ ratio due
to the small errors associated with the measurement of the strong nebular
lines of [O{\sc iii}] and the derived T$_e$([O{\sc iii}]), but increases to up
to 30\% for the O$^{+}$/H$^{+}$ ratio. The accuracies are lower in the case of
the abundances of sulphur (of the order of 25\% for S$^+$ and 15\% for
S$^{2+}$) due to the presence of larger observational errors both in the
measured line fluxes and the derived electron temperatures. The error for the
total abundance of sulphur  is also larger than in the case of oxygen (between
15\% and 30\%) due to the uncertainties in the ionization correction factors.

This is in contrast with the unrealistically small errors quoted for line
temperatures other than T$_e$([O{\sc iii}]) in the literature, in part due to
the commonly assumed methodology of deriving them from the measured
T$_e$([O{\sc iii}]) through a theoretical relation and calculating the errors
simply by propagating statistically the T$_e$([O{\sc iii}]) ones. These
relations are found  
from photo-ionization model sequences and no uncertainty is attached to them
although large scatter is found when observed values are plotted; usually the
line temperatures obtained in this way  carry only the observational error
found for the T$_e$([O{\sc iii}]) measurement and does not include the
observed scatter, thus heavily underestimating the errors in the derived
temperature.

In fact, no clear relation is found between T$_e$([O{\sc iii}]) and
T$_e$([O{\sc ii}]) for the existing sample of objects. A comparison between
measured and model derived T$_e$([O{\sc ii}]) 
shows than, in general, model predictions overestimate this temperature and
hence underestimate the O$^+$/H$^+$ ratio. This, though not very important for
high excitation objects, could be of some concern for lower excitation ones
for which total O/H abundances could be underestimated by up to 0.2 dex. It is 
worth noting that the objects observed with double-arm spectrographs,
therefore implying simultaneous and spatially coincident observations over the
whole spectral range, show less scatter in the T$_e$([O{\sc
iii}])\,-\,T$_e$([O{\sc ii}] plane clustering around the N$_e$\,=\,100
cm$^{-3}$ photo-ionization model sequence. On the other hand, this small
scatter could partially be due to the small range of temperatures shown by
these objects due to possible selection effects. This small temperature range
does not allow either to investigate the metallicity effects found in the
relations between the various line temperatures in recent photo-ionization
models by \cite{2006A&A...448..955I}.  

Also, the observed objects, though showing  Ne/O
and Ar/O relative abundances typical of those found for a large \HII\ galaxy
sample \cite{2007MNRAS.381..125P}, show higher
than typical N/O abundance 
ratios that would be even higher if the [O{\sc ii}] temperatures would be
found from photo-ionization models. We therefore conclude that the approach of
deriving the O$^+$ temperature from the O$^{2+}$ one should be discouraged if
an accurate abundance derivation is sought. 

These issues could be addressed by re-observing the objects
listed in Table 13 of \cite{2008MNRAS.383..209H}, which cover an ample range
in temperatures and metal content, 
with double arm spectrographs. This sample should be further extended  to
obtain a self consistent sample of about 50 objects with high signal-to-noise
ratio and 
excellent spectrophotometry covering simultaneously from 3600 to 9900\,\AA.
This simple and easily feasible project would provide important scientific
return in the form of critical tests of photo-ionization models. 

For the WHT objects, we have compared our obtained spectra with those
downloaded from the SDSS DR3 finding a satisfactory agreement. The analysis of
these spectra yields values of line temperatures and elemental ionic and total
abundances which are in general agreement with those derived from the WHT
spectra, although for most quantities, they can only be taken as estimates
since, due to the lack of direct measurements of the required lines,
theoretical models had to be used whose uncertainties are impossible to
quantify. Unfortunately, the spectral coverage of SDSS precludes the
simultaneous observation of the [O{\sc ii}]\,$\lambda\lambda$\,3727,29\,\AA\
and [S{\sc iii}]\,$\lambda\lambda$\,9069, 9532\,\AA\ lines, and therefore the
analysis can never be complete.

The ionization structure found for the observed objects from the 
O$^{+} $/O$^{2+} $ and S$^{+} $/S$ ^{2+} $ ratios for all the observed
galaxies, except one,  cluster around a value of the ``softness parameter"
$\eta$ of 1 implying high values of the stellar ionizing temperature. For the
discrepant object, showing a much lower value of $\eta$,  the intensity of the
[O{\sc ii}]\,$\lambda\lambda$\,7319,25\,\AA\ lines are affected by atmospheric
absorption lines. When the observational counterpart of the ionic ratios is
used, this object shows a ionization structure similar to the rest of the
observed ones. This simple exercise shows the potential of checking for
consistency in both the $\eta$ and $\eta$' plots in order to test if a given
assumed ionization structure is adequate. In fact, these consistency checks
show that the stellar ionizing temperatures found for the observed \HII\
galaxies using the ionization structure predicted by state of the art
ionization models result too low when compared to those implied by the
corresponding observed emission line ratios.  Therefore, metallicity 
calibrations based on abundances derived according to this conventional method 
are probably bound to provide metallicities which are systematically too high
and should be revised.

Finally, we have measured the Balmer continuum temperature for the three WHT
objects and derived the temperature fluctuations as defined by
\cite{1967ApJ...150..825P}. Only for one of the objects, the
temperature fluctuation is significant and could lead to higher oxygen
abundances by about 0.20\,dex.


\section*{References}
\begin{thebibliography}{10}


\bibitem{1988MNRAS.231...57D} D\'iaz, A.~I.\ 1988, \mnras, 231, 57 

\bibitem{1980ApJ...240...41F} French, H.~B.\ 1980, \apj, 240, 41 

\bibitem{2008MNRAS.383..209H} H{\"a}gele, G.~F., 
D{\'{\i}}az, {\'A}.~I., Terlevich, E., Terlevich, R., P{\'e}rez-Montero, 
E., \& Cardaci, M.~V.\ 2008, \mnras, 383, 209 

\bibitem{2006MNRAS.372..293H} H{\"a}gele, G.~F., 
P{\'e}rez-Montero, E., D{\'{\i}}az, {\'A}.~I., Terlevich, E., 
\& Terlevich, R.\ 2006, \mnras, 372, 293 

\bibitem{2006MNRAS.365..454H} Hoyos, C., \& D{\'{\i}}az, A.~I.\ 2006, \mnras,
  365, 454  

\bibitem{2006A&A...448..955I} Izotov, Y.~I., Stasi{\'n}ska, G., Meynet, G.,
  Guseva, N.~G., \& Thuan, T.~X.\ 2006, \aap, 448, 955  

\bibitem{2000A&ARv..10....1K} Kunth, D., \"Ostlin, G.\ 2000, \aapr, 10, 1 

\bibitem{1985MNRAS.216..255M} Melnick, J., Terlevich, 
R., \& Eggleton, P.~P.\ 1985, \mnras, 216, 255 

\bibitem{1985RMxAA..11...91M} Melnick, J., Terlevich, 
R., \& Moles, M.\ 1985, Revista Mexicana de Astronomia y Astrofisica, 11, 91 

\bibitem{1967ApJ...150..825P} Peimbert, M.\ 1967, \apj, 
150, 825 

\bibitem{2005MNRAS.361.1063P} P{\'e}rez-Montero, E., \& D{\'{\i}}az, A.~I.\
  2005, \mnras, 361, 1063  

\bibitem{2007MNRAS.381..125P} 
P{\'e}rez-Montero, E., H{\"a}gele, G.~F., Contini, T., 
\& D{\'{\i}}az, {\'A}.~I.\ 2007, \mnras, 381, 125 

\bibitem{1970ApJ...162L.155S} Sargent, W.~L.~W., \& Searle, L.\ 1970, \apjl,
  162, L155 

\bibitem{1998ApJ...493L..23S} Schulte-Ladbeck, R.~E., Crone, M.~M., \& Hopp,
  U.\ 1998, \apjl, 493, L23  

\bibitem{2002AJ....123..485S} Stoughton, C., et 
al.\ 2002, \aj, 123, 485 

\bibitem{1991A&AS...91..285T} Terlevich, R., Melnick, J., Masegosa, J., Moles,
  M., \& Copetti, M.~V.~F.\ 1991, \aaps, 91, 285  







\end{thebibliography}
\end{document}